\documentclass[12pt]{article}
\usepackage{graphicx}
\input epsf
\usepackage{amssymb}
\parskip=\medskipamount
\def\al{\alpha}
\def\be{\beta}

\def\om{\omega}   
\def\ID{\relax{\rm l\kern-.18 em D}}
\def\IE{\relax{\rm l\kern-.18 em E}}
\def\IK{\relax{\rm l\kern-.18 em K}}
\def\IL{\relax{\rm I\kern-.18 em L}}
\def\IN{\relax{\rm I\kern-.18 em N}}
\def\IR{\relax{\rm I\kern-.18 em R}}
      
\def\uno{\relax{\rm 1\kern-.18 em l}}

\def\IK{\relax{\rm l\kern-.18 em K}}
\def\IL{\relax{\rm I\kern-.18 em L}}
\def\IN{{\Bbb N}}
\def\IR{{\Bbb  R}}

\def\Re{\mathop{\rm Re}\nolimits}
\def\Im{\mathop{\rm Im}\nolimits}

\def\smallonehalf{\frac{{}_1}{{}^2}}

\def\wt{\widetilde}
\def\frac#1#2{{#1\over #2}}

\def\ptos{\leaders\hbox to 2mm{\hfil{.}\hfil}\hfill}
\def\\{\hfill\break}

\def\<#1>{\langle#1\rangle}

\font\tenfrak=eufm10  \font\sevenfrak=eufm7  \font\fivefrak=eufm5
\newfam\frakfam
\textfont\frakfam=\tenfrak\scriptfont\frakfam=\sevenfrak
\scriptscriptfont\frakfam=\fivefrak

\font\tengoth=eufm10 scaled\magstep1 \font\sevengoth=eufm7
\font\fivegoth=eufm5
\newfam\gothfam
\textfont\gothfam=\tengoth\scriptfont\gothfam=\sevengoth
   \scriptscriptfont\gothfam=\fivegoth

\newtheorem{proposicion}{Proposition}

\begin{document}

\title{Higher-order superintegrability  of separable potentials with a new approach to  the  Post-Winternitz system }

\author{ Manuel F. Ra\~nada  \\ [3pt]
 {\sl Departamento de F\'{\i}sica Te\'orica  and IUMA} \\
 {\sl Universidad de Zaragoza, 50009 Zaragoza, Spain}  }
\date{}
\maketitle

\begin{abstract}  
The higher-order superintegrability of separable potentials is studied. 
It is proved that these potentials possess (in addition to the two quadratic integrals) a third integral of higher-order in the momenta that can be obtained as the product of  powers  of two particular rather simple complex functions.   Some systems related with the harmonic oscillator, as the generalized SW system and the TTW system, were studied in previous papers; now a similar analysis is presented for superintegrable systems related with the Kepler problem.  In this way, a new proof of   the superintegrability of the Post-Winternitz system is presented and the explicit expression of the integral is obtained. 
Finally, the relations between the superintegrable systems with quadratic constants of the motion (separable in several different coordinate systems) and the superintegrable systems  with higher-order constants of the motion are analyzed. 
\end{abstract}

\begin{quote}
Higher-order  constants of motion. Complex factorization. 

{\sl Running title:}{\enskip}
Higher-order superintegrability  of separable potentials.



\end{quote}

\vfill
\footnoterule{\small
\begin{quote}
{\tt E-mail: {mfran@unizar.es}  }
\end{quote}
}

\newpage

\section{Introduction}

It is known that there are three different classes of superintegrable systems : 
(i)  Superseparable, (ii) Separable, and (iii) Nonseparable.
\begin{itemize}
\item[(i)] Systems that admit  separability in at least two different coordinate systems  are superintegrable  with constants of motion  linear or quadratic in the momenta. 
In the Euclidean plane, Fris {\sl et al. }\cite{FrMS65}  proved the existence of four families of  potentials separable in two different sets of coordinates
\begin{enumerate}
\item[(1)] Two systems related with the harmonic oscillator  
\begin{eqnarray}  
   V_{a}  &=&  {\smallonehalf}\, \om_0^2(x^2 + y^2)  + \frac{k_1}{x^2}  
   + \frac{k_2}{y^2} \,, \\
   V_{b}  &=&  {\smallonehalf}\, \om_0^2(x^2 + 4y^2)  + \frac{k_1}{x^2}  
   + k_2 y\,. 
\end{eqnarray}   
$V_a$ is separable in Cartesian and polar  and $V_b$ in Cartesian and parabolic coordinates. 
\item[(2)] Two systems related with the Kepler problem 
\begin{eqnarray}  
   V_{c} &=& \frac{g}{r} + \frac{k_1}{y^2}  + \frac{k_2}{r}\frac{x}{y^2} \,,\\
   V_{d} &=& \frac{g}{r} + k_1\,\frac{\sqrt{r + x}}{ r}  
   + k_2\,\frac{\sqrt{r - x}}{ r}    \,. 
\end{eqnarray}  
$V_c$ is separable in polar and   parabolic and $V_d$ in two different systems of parabolic coordinates. 
\end{enumerate}
The properties of these four systems have been studied for many authors from both  the classical  (see e.g. \cite{Ev90}--\cite{KaKrPM01}  and references therein) and the quantum  \cite{ShTW01}--\cite{KaWiMiPo02} viewpoint.

\item[(ii)]  Separable systems in only one coordinate system. 

  If the system has $n$ degrees of freedom then it possess a first set of $n$ quadratic  constants of motion (determined by the separability) and a second set of $n-1$ constants of higher order than two \cite{GrW02}-\cite{PoPoW12}.

\item[(iii)]  Nonseparable systems. 

Very few systems of this class are known. We can mention a very particular case of the Holt potential \cite{PostWint11} and the  Calogero-Moser system \cite{Woj83}--\cite{SmW06} with constants of the motion related with  a Lax pair of matrices.  We can also include in this set systems with nonstandard Lagrangians as a nonlinear oscillator related with the Riccati equation \cite{CaRS05}.  
\end{itemize}

We must underline that we are considering natural Hamiltonians (that is, kinetic term plus a potential).  In fact, the concept of superintegrability has also been  extended to the case of Hamiltonians describing systems in magnetic fields (velocity-dependent terms in the potential) \cite{McSWint00}-\cite{Mar11}
and to quantum systems which include spin interactions \cite{WinterYur06}-\cite{Nik12}. We note that if we consider velocity-dependent potentials then quadratic integrability no longer implies the separation of variables and the constants of the motion have both even and odd powers. These two classes of systems are not considered in this paper.  

We mention now two recently discovered systems belonging to the class (ii) of superintegrable systems that are separable  in only one coordinate system.   

\begin{enumerate}
\item[(1)]  Tremblay, Turbiner, and Winternitz \cite{TTW09}--\cite{TTW10}, and then other authors  \cite{Qu10a}--\cite{Ra12b},  have studied the following $k$-dependent family of potentials 
$$
  V_{ttw}(r,\phi)  =  {\smallonehalf}\,{\om_0}^2 r^2 +  \frac{1}{2\,r^2}\,\Bigl(   \frac{\alpha} {\cos^2(k\phi)} +  \frac {\beta} {\sin^2(k\phi)} \Bigr) \,.
  $$
When $k=1$ it becomes the $V_a$ potential but in the general $k\ne 1$ case  it  is only separable in polar coordinates. An important point is that the third integral,  that is a polynomial in the momenta of higher order than two, has been explicitly obtained in \cite{Ra12b} as the product of two particular rather simple complex functions. 

\item[(2)]   There is another interesting system  introduced by Post and Winternitz \cite{PostWint10} that has certain similarity with  the TTW system but that is related, not with the harmonic oscillator, but with the Kepler problem (hydrogen atom in the quantum case).  These two authors studied this system making use of  the so-called coupling constant  metamorphosis  (St\"ackel transform) \cite{KaMiPost10} as an approach. 
\end{enumerate}

We studied in Ref.  \cite{Ra12b} the higher-order superintegrability of separable systems related with the harmonic oscillator and we proved the superintegrability of the TTW system. The purpose of this paper is threefold: 
\begin{itemize}
\item  To present a similar analysis but of systems related with the Kepler problem. 
\item  To present a new proof of the superintegrability of the Post-Winternitz system. 
\item  To study in depth the relations between the superintegrable systems of type (i) and of type (ii). 
\end{itemize}

In more detail, the structure of the article is as follows:
In Sec. 2 we study the higher-order superintegrability of two systems related with  the harmonic oscillator and we obtain the third constant of the motion as the product of powers  of two particular rather simple complex functions. Then in Sec. 3 
we apply the formalism studied in Sec. 2 to new family of superintegrable potentials related with the Kepler problem and we prove the superintagrability of the Post-Winternitz system. 
Finally  in Sec. 4 we make some comments, present some open questions, and we finalize with a final conjecture.

\section{Two superintegrable systems related with the harmonic oscillator }

First a comment on an interesting property of the potential $V_a$. It is separable in both Cartesian and polar coordinates and this means  the existence  of two quadratic constants of motion with quadratic terms of the form $p_x^2$ (or $p_y^2$) and $(x p_y - y p_x)^2$. On the other hand, constants of the motion with  a quadratic term of the form $a p_x^2 + b (x p_y - y p_x)^2$ are associated to elliptic coordinates ($a$ and $b$ are coefficients related with the caracteristics of the conics). Hence $V_a$ is also separable in 
elliptic coordinates. But this new separability can be considered as a consequence of the two other separabilities. If we modify $V_a$ detroying Cartesian (or polar) separability then the elliptic separability is also destroyed.

In the following, we will make use of the Hamiltonian formalism; therefore,  the time derivative $d/dt$ of a function means the Poisson bracket of the function with the Hamiltonian.

\subsection{A nonlinear oscillator, with higher-order integrals of motion, separable in Cartesian coordinates } \label{Sec21}

The following potential  
\begin{equation}
  V_{a}(n_x,n_y)  =  {\smallonehalf}{\om_0}^2(n_x^2 x^2 + n_y^2 y^2)
    + \frac{k_1}{2x^2}  + \frac{k_2}{2y^2}  \,,  \label{Van1n2}
\end{equation}  
(the ratio $n_y/n_x$ is a rational number) reduces to $V_a$ when $n_x=n_y=1$ but in the general case it is only separable in Cartesian coordinates. It has been proved that it is superintegrable \cite{EvVe08}--\cite{Ra12a} with  a polynomial in the momenta of higher order than two as a  third integral.  

 When $k_1=k_2=0$ this potential reduces  to the harmonic oscillator and then the third integral can be factorized \cite{JauHillPr40,Perelomov} as the product  of powers of the following functions
$$
 A_x = p_x + i\, n_x {\om_0}\, x \,,{\quad}
 A_y = p_y + i\, n_y {\om_0}\, y \,.   
$$
It has been proved in \cite{RaRoS10} that, in the general case, $k_1\ne 0$, $k_2\ne 0$, the third integral can also be factorized as the product of powers of  two complex functions. If we denote by  $B_i$, $i=x,y$, the functions 
$$
 B_{x} = A_x^2 + \frac{k_1}{x^2} \,,{\quad}
 B_{y} = A_y^2 + \frac{k_2}{y^2}  \,,  
$$
then we have
$$
 \frac{d}{d t}\,B_x = 2\,i\, n_x {\om_0}\,B_x  \,,{\quad}
 \frac{d}{d t}\,B_y = 2\,i\, n_y {\om_0}\,B_y  \,.
$$
Thus,  the complex functions $B_{ij}$ given by 
$$
  B_{ij} = (B_i)^{n_j}\,(B_j^{*})^{n_i}\,,{\quad} i,j=x,y,
$$
are constants of the motion
$$
  \frac{d}{dt}\,B_{xy}  =   B_x^{(n_y-1)}B_y^{(n_x-1)}\,\Bigl( \, 
  n_y\dot{B_x}\,B_y^{*}  +  n_x B_x\,\dot{B}_y^{*} \,\Bigr)  =  0  \,.
$$

The two real functions $|B_{xx}|^2$ and $|B_{yy}|^2$ are related with the 
two one-dimensional energies $I_1=E_x$ and $I_2=E_y$  and
concerning $B_{xy}$, as it is a complex function, it determines
not just one   but two real first integrals, $\Re (B_{xy})$ and $\Im
(B_{xy})$.   We can choose  one of them as the third fundamental constant   of motion 
(the other constant $I_4=\Re(A_{xy})$ can be expressed as a function 
of $E_x$, $E_y$, and $\Im(A_{xy})$).

\subsection{A nonlinear oscillator, with higher-order integrals of motion, separable in Cartesian coordinates II} \label{Sec22}

The following potential  
\begin{equation}
  V_{b}(n_x,n_y)  =  {\smallonehalf}{\om_0}^2(n_x^2 x^2 + n_y^2 y^2)
    + \frac{k_1}{2x^2}  + k_2 y  \,,  \label{Vbn1n2}
\end{equation}  
reduces to $V_b$ when $n_x=1, n_y=2$ (the $k_2$ dependent term can be obtained from the $y^2$ term by a translation). The potential $V_b$ is separable in Cartesian and parabolic coordinates but in the general $(n_x,n_y)$ case only Cartesian separability is preserved. It is superintegrable with a third constant of the motion given by 
$$
  C_{xy} = (B_x)^{n_y}\,(\wt{A}_y^{*})^{2n_x}\,,{\quad} \wt{A}_y=p_y + i\,  (n_y {\om_0}\, y +    k_2' ) \,,  
$$
with $k_2'=k_2/n_y\om_0$. It can be verified that in the particular $n_x=1, n_y=2$ case the imaginary part of $C_{xy}$ leads (after some simplification) to a quadratic constant of the motion given by 
$$
 I_3(1,2) = (x p_y - y p_x) p_x + {\om_0}^2 x^2 y - k_1 \frac{y}{x^2}
$$
that is the integral of motion arising from the separability in parabolic coordinates.

\subsection{A nonlinear oscillator, with higher-order integrals of motion, separable  in polar coordinates } \label{Sec23}

It  was proved in Ref. \cite{Ra12b} that the following Hamiltonian 
\begin{equation}
 H_{ak} =   {\smallonehalf}\,\Bigl(p_r^2 + \frac{p_\phi^2}{r^2}\Bigr) + V_{ak}    \,,{\quad}
 V_{ak} = {\smallonehalf}\,{\om_0}^2 r^2 + {\smallonehalf}\frac{F(\phi)}{r^2} \,,  
 \label{H(rfi)}
\end{equation}
that is separable in polar coordinates with the following  two constants of the motion 
\begin{eqnarray*}  
 J_1 &=& p_r^2 + \frac{p_\phi^2}{r^2} + {\om_0}^2 r^2 +\frac{F(\phi)}{r^2} \,, \cr 
 J_2 &=& p_\phi^2 + F(\phi)  \,, 
\end{eqnarray*}
it is also superintegrable with a third constant of the motion of higher order in the momenta when the function $F$ takes the form of a $k$-dependent function $F_k$  given by 
\begin{equation}
  F_{k} = \frac{k_a} {\sin^2(k\phi)} +  k_b\,\Bigl(\frac {\cos(k\phi)} {\sin^2(k\phi)}\Bigr) \,,
\end{equation}
where $k_a$ and $k_b$ are arbitrary constants. 

The proof is as follows: 
Let $M_r$ and $N_\phi$ be the  complex functions   $M_r = M_1 + i\,M_2$ and  $N_\phi = N_1 + i\,N_2$ with real and imaginary parts $M_a$ and $N_a$, $a=1,2$, be defined as 
$$
 M_1 =  \frac{2}{r}\,p_r\,\sqrt{J_2} \,,{\quad}
 M_2 =   p_r^2 + {\om_0}^2 r^2 -  \frac{J_2}{r^2} \,   \,,
$$
$$
 N_1 =  \frac{k_b}{2} +  J_2\,\cos(k\phi)  \,,{\quad}
 N_2 =   \sqrt{J_2} \,p_\phi\,\sin(k\phi) \,.
$$ 
Then, time-evolution of the functions $M$ and $N$ is given by
$$
 \frac{d}{d t}\,M_r  = i\, 2 {\lambda}\,M_r  \,,{\quad}
 \frac{d}{d t}\,N_\phi  = i\, k\,{\lambda}\,N_\phi   \,,{\quad}
  {\lambda} = \frac{1}{r^2}\,\sqrt{J_2}\,. 
$$
Therefore the complex function $K_k$ defined as
\begin{equation}
  K_k = M_r^{k} \,(N_\phi^{*})^2   \label{Kk}
\end{equation}
is a (complex) constant of the motion.

The angular function $F_k(\phi)$, that was obtained in  \cite{Ra12b} as the solution of a linear differential equation (related with the function $N_\phi$), can alternatively be presented in some other forms. In fact,  it can be proved that the function $F_{k}$ reduces to the $\phi$-dependent part of the TTW potential 
\begin{equation}
  V_{ttw}(r,\phi)  =  {\smallonehalf}\,{\om_0}^2 r^2 +  \frac{1}{2\,r^2}\,\Bigl(   \frac{\alpha} {\cos^2(k\phi)} +  \frac {\beta} {\sin^2(k\phi)} \Bigr) \,,
\end{equation}
when (i) the coefficients $k_a$ and $k_b$ are writen as $k_a=2(\alpha+\beta)$ and $k_b=2(\beta -\alpha)$, and (ii) the parameter $k$ is changed to $2k$; that is,  
$$
  \frac{2(\alpha+\beta)} {\sin^2(2k\phi)} +  2(\beta - \alpha)\,\Bigl(\frac {\cos(2 k\phi)} {\sin^2(2k\phi)}\Bigr) =  \frac{\alpha} {\cos^2(k\phi)} +  \frac {\beta} {\sin^2(k\phi)}  \,.   
$$
Consequently we have 
$$
  V_{ak}(r,\phi,2k)  =  V_{ttw}(r,\phi,k)  \,,  
$$
and the above function $K_k$  also represents (introducing the appropriate changes) the third constant of the motion of the TTW system.

\section{A superintegrable system related with the Kepler problem}\label{Sec3}

In the factorization (\ref{Kk}) of the function $K_k$, the factor $N_\phi$ is a function depending only on the angular variables;  a consequence of this is that the property 
$$
 \frac{d}{d t}\,N_1 =  (-{\lambda})\,k\,N_2  \,,{\quad} 
 \frac{d}{d t}\,N_2  =  {\lambda}\,k\,N_1 \,,
$$
remains true for any potential of the form 
\begin{equation}
  V(r,\phi)  = U(r) + {\smallonehalf}\frac{F_{k}(\phi)}{r^2}    \,.  \label{UFk1}
\end{equation}
Thus, if we search for an appropriate factorization of a third constant of the motion for a separable potential as  (\ref{UFk1}), but with the function $U(r)$ representing the Kepler potential,  then we must preserve $N_\phi$ and look only for a new function $M_r$.

 The following proposition states the result obtained for the existence of a superintegrable system $V_{ck}$ that generalizes the potential $V_c$.
 
\begin{proposicion}  \label{prop1}
Consider the following Kepler-related  potential
\begin{equation}
  V_{ck}(r,\phi)  =  -\frac{g}{r} + {\smallonehalf}\frac{F_{k} (\phi)}{r^2}   \,,{\quad}
  F_{k}  = \frac{k_a} {\sin^2(k\phi)} +  k_b\,\Bigl(\frac {\cos(k\phi)} {\sin^2(k\phi)}\Bigr) \,,
\end{equation}
where $k_a$ and $k_b$ are arbitrary constants, and let us denote by $J_1$ and $J_2$ the two constants of the motion associated to the separability in polar coordinates 
\begin{eqnarray*}  
 J_1 &=& p_r^2 + \frac{p_\phi^2}{r^2}  -\frac{2g}{r}  +\frac{F_{k}(\phi)}{r^2}  \,,\cr 
 J_2 &=& p_\phi^2 + F_{k}(\phi)  \,.
\end{eqnarray*}
Let $M_r$ and $N_\phi$ be the  complex functions   $M_r = M_1 + i\,M_2$ and  $N_\phi = N_1 + i\,N_2$ with real and imaginary parts $M_a$ and $N_a$, $a=1,2$,  given by 
$$
 M_1 =  p_r\,\sqrt{J_2} \,,{\quad}
 M_2 =  g -  \frac{J_2}{r}   \,,
$$
$$
 N_1 =  \frac{k_b}{2} +  J_2\,\cos(k\phi)   \,,{\quad}
 N_2 =  \sqrt{J_2} \,p_\phi\,\sin(k\phi) \,. 
$$ 
Then, the complex function $K_k$ defined as 
$$
  K_k = M_r^{k} \,N_\phi^{*} 
$$
is a (complex) constant of the motion.
\end{proposicion}

{\noindent}{\bf Proof:} 
The time-evolution of the functions $M_r$ and $N_\phi$ is given by
$$
 \frac{d}{d t}\,M_r  =   i\,  {\lambda}\,M_r  \,,{\quad}
 \frac{d}{d t}\,N_\phi  =  i\, k\,{\lambda}\,N_\phi   \,,{\quad}
  {\lambda} = \frac{1}{r^2}\,\sqrt{J_2}\,. 
$$
Thus we have
$$
  \frac{d}{dt}\,(M_r^{k} \,N_\phi^{*} )  =   M_r^{(k-1)}\,\Bigl(  k\,\dot{M_r}\,N_\phi^{*}
   +   M_r\,\dot{N_\phi}^{*} \Bigr)  =  0   \,. 
$$
\hfill$\square$

The function $F_k$ in $V_{ck}$ is the same as in $V_{ak}$; so if we denote by $V_{pw}(r,\phi) $ the potential studied by Post and Winternitz in Ref. \cite{PostWint10}
\begin{equation}
  V_{pw}(r,\phi)  =   -\frac{g}{r}  +  \frac{1}{2\,r^2}\,\Bigl(   \frac{\alpha} {\cos^2(k\phi)} +  \frac {\beta} {\sin^2(k\phi)} \Bigr) \,,
\end{equation}
then we have
$$
  V_{ck}(r,\phi,2k)  =  V_{pw}(r,\phi,k)  \,,  
$$
and the above function $K_k$  also represents (introducing the appropriate changes) the third constant of the motion of the PW system.

We close this section with the following two poits: 
\begin{itemize}

 \item    The expressions of the function $F_{k}$ for the three first integer values of $k$ are :
\begin{eqnarray*}
 F_{1} &=& \frac{k_a}{y^2} + \frac{k_k\,x}{y^2\sqrt{x^2+y^2}}  
 \,,{\quad} (k=1)  \cr
 F_{2} &=& \frac{k_a-k_b}{4x^2} + \frac{k_a+k_b}{4y^2}  
  \,,{\quad} (k=2)  \cr
 F_{3}  &=& \frac{1}{(3x^2-y^2)^2y^2} \Bigl( k_a (x^2+y^2)^2 
 + k_b (x^2-3y^2) x \sqrt{x^2+y^2}  \Bigr) 
   \,,{\quad} (k=3)
\end{eqnarray*}
\item The interchange the  functions $\cos(k\phi)$ and $\sin(k\phi)$ leads to a new function $G_k$ given by 
$$
  G_{k} = \frac{k_a} {\cos^2(k\phi)} +  k_b\,\Bigl(\frac {\sin(k\phi)} {\cos^2(k\phi)}\Bigr)\,,    
$$
that determines a new (but related) superintegrable system  
\begin{equation}
 H =   {\smallonehalf}\,\Bigl(p_r^2 + \frac{p_\phi^2}{r^2}\Bigr) + V_{ck}' 
 \,,{\quad}
 V_{ck}' = -\frac{g}{r} +  {\smallonehalf}\frac{G_{k}(\phi)}{r^2} \,.  
 \label{H(rfi)}
\end{equation}

This interchange of functions corresponds to a rotation of angle $\pi/(2 k)$, that is,   $V_{ck}'(r,\phi)  = V_{ck}\bigl(r,\phi + \pi/(2 k)\bigr)$. 
Therefore, the potential  $V_{ck}'$ can be seen not just as a new potential but as a rotated version of $V_{ck}$. Nevertheless  the rotation is different  for every value of $k$ and in some cases (noninteger values of $k$) the domain of $V_{ck}'$ can be different from the domain of $V_{ck}$.

\end{itemize}

\section{Some comments with a final conjecture  } 

This paper is mainly concerned with the study of the following two points.

\begin{itemize}
\item[(i)]  Existence of families of separable systems possessing a higher constant of the motion (belonging therefore to the class (ii)), that can be considered as generalizations or deformations of the potentials $V_a$, $V_b$, and $V_c$. 

\item[(ii)]  Proof that the third integral of the motion can be expressed, in the four cases,  as the product of powers  of two particular rather simple complex functions (denoted by $A_i$, $B_i$, $\wt{A}_i$, $M_a$ and $N_a$). 
\end{itemize}

The point (i) can be considered as a prolongation of previous studies \cite{Ra12b}, \cite{RaRoS10}, on the TTW  \cite{TTW09}--\cite{Ra12b}, and  on the generalized SW  \cite{EvVe08} --\cite{Ra12a} potentials  (see also \cite{LevPoWint12}   for a  related study in the pseudo-Euclidean plane). The point (ii) is certainly important since  the existence of this factorization provides a direct and elegant method for obtaining the additional integral.

 We have pointed out in the Introduction that all the quadratic Euclidean superintegrable systems (belonging to the class (i)) can be considered as deformations of the  harmonic oscillator or of the Kepler potential  with the coeficients $k_1$ and $k_2$ representing the intensity of the deformation (we recall that we are considering Hamiltonians with scalar or natural potentials and hence excluding other more general forms of potentials).   Now we see that this property seems to be also true for these more general families with superintegrability of type (ii). 

 The actual situation can be summarized as follows.
 
\begin{enumerate}
\item[(a)]  There exist two different families of separable potentials generalizing the potential $V_a$. This means that  
\begin{enumerate}
\item[(a1)]  $V_a$ appears as the $V_a(1,1)$ particular case of the family  $V_{a}(n_x,n_y)$  that preserves Cartesian but not polar separability. 
\item[(a2)]  $V_a$ appears as the $k=2$ particular case of the family $V_{ak}(r,\phi)$ [or as the $k=1$ case of $V_{ttw}(r,\phi)$]    that preserves polar but not Cartesian separability. 
\end{enumerate}

\item[(b)]  There exist a family of separable potentials generalizing the potential $V_b$. That is, $V_b$ appears as the $V_b(1,2)$ particular case of the family  $V_{b}(n_x,n_y)$ that preserves Cartesian but not parabolic separability. 

\item[(c)]  There exist a family of separable potentials generalizing the potential $V_c$. That is, $V_c$ appears as the $k=1$ particular case of the family $V_{ck}(r,\phi)$ [or as the $k=1/2$ case of $V_{pw}(r,\phi)$] that preserves polar but not parabolic separability. 
\end{enumerate}

At this point we recall the expressions of the potentials $V_{b}$, $V_{c}$, and $V_{d}$, when written in parabolic coordinates
\begin{eqnarray*}  
   V_{b} &=&     {\smallonehalf}\, \om_0^2 \,(\al^4 - \al^2 \be^2 + \be^4)  +  \frac{k_1}{\al^2\,\be^2}  + k_2 \, (\al^2 - \be^2 )    \,, \\
   V_{c}   &=&    \frac{g}{\al^2 + \be^2}  + \,\frac{k_1}{\al^2\,\be^2}   +  \frac{k_2}{\al^2  \be^2} \frac{\be^2 - \al^2}{\al^2 + \be^2}     \,, \\
   V_{d}   &=&     \frac{g}{\al^2 + \be^2} +   \frac{k_1\,\al}{\al^2 + \be^2} +   \frac{k_2\,\be}{\al^2 + \be^2} \,. 
\end{eqnarray*}  

It seems natural to formulate the following two points as a conjecture.
\begin{enumerate}
\item[(1)]   There exists two other families of higher order superintegrable systems  that are separable in parabolic coordinates and reduce, when $k_1=k_2=0$, to the harmonic oscillator $(1/2)(x^2 + 4 y^2)$ (with ratio of frequencies $n_y/n_x=2/1$) in one case and to the Kepler potential in the other case. The oscillator family must be related with $V_b$ and the Kepler family with $V_c$ or/and $V_d$. 

\item[(2)]   These two families with parabolic separability  are superintegrable with a third integral of the motion that  can also be expressed as the product of powers of two particular rather simple complex functions. 
\end{enumerate}

For the moment these two points must be considered as suppositions to be proved.   These questions, as well as some other related problems as for example,    existence of bi-Hamiltonian structures  related with the complex factorization (this was done in \cite{CaMR02} for the harmonic oscillator) or application of the complex factorization (changing functions for opertors) to the resolution of the associated quantum Schr\"odinger equation, are open questions to be studied.

\section*{Acknowledgments}

Discussions with M. Santander on the properties of superintegrable systems are gratefully acknowledged. 
This work was supported by the research projects MTM--2009--11154 (MEC, Madrid)  and DGA-E24/1 (DGA, Zaragoza).

{\small

\end{document}